\documentclass{article}
\usepackage{spconf,amsmath,amssymb,graphicx}

\title{Study of Multi-Branch Tomlinson-Harashima Precoding with Multiple-Antenna Systems and  Rate Splitting}
\name{A. R. Flores$^{1}$, R. C. de Lamare$^{1,2}$ and B. Clerckx$^{3}$ \thanks{Thanks to CNPq agency for funding.}}
\address{
$^{1}$CETUC, Pontifical Catholic University of Rio de Janeiro, Brazil\\
$^{2}$Department of Electronics, University of York, UK\\
$^{3}$Department of Electrical and Electronic Engineering, Imperial College London, London SW7
2AZ, UK\\
Emails:andre.flores@cetuc.puc-rio.br, delamare@cetuc.puc-rio.br,b.clerckx@imperial.ac.uk}
\begin{document}
\ninept
\maketitle
\begin{abstract}
Rate splitting (RS) has emerged as a valuable technology for wireless communications systems due to its capability to deal with uncertainties in the channel state information at the transmitter (CSIT). RS with linear and non-linear precoders, such as the Tomlinson-Harashima (THP) precoder, have been explored in the downlink (DL) of multiuser multi antenna systems. In this work, we propose a multi-branch (MB) scheme for a RS-based multiple-antenna system, which creates patterns to order the transmitted symbols and enhances the overall sum rate performance compared to existing approaches. Closed-form expressions are derived for the sum rate through statistical analysis. Simulation results show that the proposed MB-THP for RS outperforms conventional THP and MB-THP schemes.
\end{abstract}
\section{Introduction}
\label{sec:intro}

Over the past ten years, notable improvements in multiple-input multiple-output (MIMO) technology for wireless communications systems have been achieved. The main advantage of MIMO is that it increases the overall throughput of the system without needing additional bandwidth. For this reason, MIMO has been contemplated in diverse communications standards, such as IEEE802.11n, 3 GPP long-term evolution (LTE) and 5G \cite{Li2010,Jones2015,Parkvall2017,Shafi2017,mmimo,wence}. Nevertheless, MIMO systems are affected by multiuser interference (MUI), which degrades and limits the overall performance of the system.

To cope with the MUI, receive processing techniques in the uplink (UL) and transmit processing techniques in the downlink of MIMO system have been developed. The transmit processing usually involves the design of a precoder, which maps the symbols 
4
 to the transmit antennas. In this sense both linear and non-linear precoding techniques \cite{joham2005,Tomlinson1971,Harashima1972,Sung2009,gbd,wlbd,Zu2014,rthp,rmmse,bbprec,baplnc,lrcc,mwmimo,plszf,rmmsecf} have been reported in literature, where it is well known that non-linear precoders outperform their linear counterparts. Unfortunately, the design of the precoder requires the perfect knowledge of the channel state information  at the transmitter (CSIT). In practice, this task is extremely difficult to perform as only imperfect or partial CSIT is available, which creates residual MUI. Therefore, the perfect CSIT assumption remains questionable. 

In these circunstamces, Rate Splitting (RS) has raised as an attractive solution to deal with inaccurate CSIT \cite{Clerckx2016}. RS splits the message of each user into two different parts known as common message and private message. The common messages of all users are combined and encoded in one common stream, while the private message of each user is encoded into a private stream. At the receiver, the common stream is decoded first. Afterwards, successive interference cancellation (SIC) is performed at each user to retrieve its private stream. The common stream should be decoded by all the users, whereas the private streams are decoded only by its corresponding user. Each user reconstructs its original message by recombining the retrieved common and private message. One of the benefits of RS is that it allows us to adjust the content and the power of the common stream resulting in a more flexible and robust wireless system. 

Due to its benefits, RS has been investigated under several deployments. In all cases, RS has shown a better performance than conventional precoding strategies, Non-Orthogonal Multiple Access (NOMA) \cite{Mao2018} and even Dirty Paper Coding (DPC) \cite{Mao2020}. Linear precoding techniques with RS have been explored in \cite{Joudeh2016}, where the sum rate performance with partial CSIT is maximized . In \cite{Hao2015,Lu2018} the effects of the imperfect CSIT caused by finite feedback are mitigated. In contrast, non-linear precoders, such as the THP, have been considered in \cite{Flores2018}. The performance of the THP algorithm is strongly related to the symbol ordering. Although the previous work explores the Tomlinson-Harashima precoder (THP) along with RS, no ordering scheme over the symbols is implemented.   

In this work, we propose a multi-branch (MB) scheme for an RS MIMO system employing THP precoders, denoted as RS-MB-THP \cite{rsthp}. This technique is incorporated in two different THP deployments, namely the centralized THP (cTHP) and the descentralized THP (dTHP) architectures. The MB technique designs efficient transmit patterns to order the symbols and further enhances the performance obtained by the THP operating in an RS system. Finding the best pattern in an RS system can be computationally expensive. In this regard, suboptimal strategies are explored in order to reduce the complexity required by the evaluation of multiple patterns. Closed-form expressions to compute the SINR of the proposed schemes are also derived and numerical results illustrate the performance of the proposed RS-MB-THP against existing schemes.

This paper is organized as follows. In Section 2, the system model is described. In Section 3 the proposed RS-MB-THP is presented along with two suboptimal strategies, which reduce the computational complexity. In Section 4 the computation of the sum rate performance through closed form expressions is detailed. Section 5 presents the simulations carried out to evaluate the performance of the proposed scheme. Finally, Section 5 draws the conclusions of this work.

\section{System Model}
\label{sec:system model}

We consider the downlink (DL) of a multi-user MIMO (MU-MIMO) system where the BS serves $K$ users which are geographically distributed. Each user equipment (UE) has a single receive antenna. It follows that the total number of receive antennas is equal to $K$. In contrast, the Base Station (BS) is equipped with $N_t$ transmit antennas, where $N_t \geq K \geq 2$. The system employs an RS scheme, which splits the message of a single user. Let us consider for instance that the message, denoted by $m^{\left(RS\right)}$, of an arbitrary user $k$  is split into a common message $m_c$ and a private message $m_k$. All other messages $m_i$ with $i\neq k$ are not split. Once the splitting process is complete, the messages are encoded and modulated. Then the common symbol $s_c$ is superimposed to the vector of private symbols $\mathbf{s_p}$, i.e., $\mathbf{s}=\left[s_c,\mathbf{s}_p\right]\in \mathbb{C}^{K+1}$. Then, the vector of private symbols is given by $\mathbf{s}_p=\left[s_1,s_2,\cdots,s_K\right] \in \mathbb{C}^{K}$. 

The symbols are mapped to the transmit antennas using a precoding matrix $\mathbf{P}=\left[\mathbf{p}_c,\mathbf{p}_1,\cdots,\mathbf{p}_K\right]\in\mathbb{C}^{N_t\times\left(K+1\right)}$. Specifically, the private precoder $\mathbf{p}_k, k=1,\cdots,K$ maps the private symbol $s_k$ to the transmit antennas. On the other hand, the precoder $\mathbf{p}_c$ maps the common symbol $s_c$. Note that the system incorporates non-linear precoders, i.e., non-linear processing is performed over the data symbols creating a random vector $\mathbf{v} \in {K+1}$. Since we consider a THP precoder, a feedback filter successively cancels the interference and a modulo operation is implemented to mitigate the effects of the power loss, leading to $\mathbb{E}\left[\lVert\mathbf{s}\rVert^2\right]\approx\mathbb{E}\left[\lVert\mathbf{v}\rVert^2\right]$. However, the modulo operation introduces a modulo loss. These losses vanish as the modulation size grows. Thus, for a moderate or large constellation we have that the power of $\mathbf{v}$ is approximately equal to the power of $\mathbf{s}$ \cite{Zu2014}. Then, the transmit vector is given by
\begin{equation}
\mathbf{x}=\mathbf{P}\mathbf{v},
\end{equation}
where $\mathbf{x}\in \mathbb{C}^{N_t}$. We also consider that the system follows an average transmit power constraint given by $\mathbb{E}\left[\lVert\mathbf{x}\rVert^2\right]\leq E_{tr}$, where $E_{tr}$ denotes the available power.

The data is transmitted over a flat fading channel denoted by $\mathbf{H}^{\text{T}}=\left[\mathbf{h}_1,\cdots,\mathbf{h}_K\right]^{\text{T}}$, whose parameters remain fixed during the transmission of a data packet. The channel $\mathbf{h}_k^{\text{T}}$ represents the channel between the BS and the $k$th user terminal. Remark that in real applications there always exist uncertainty in the CSIT due the channel estimation procedure \cite{Vu2007,Love2008}. This uncertainty can be modelled by an error channel matrix given by $\mathbf{\tilde{H}}^{\text{T}}=\left[\mathbf{\tilde{h}}_1,\cdots,\mathbf{\tilde{h}}_K\right]^{\text{T}}\in \mathbb{C}^{N_t\times{K}}$. The channel estimate is expressed by $\mathbf{\hat{H}}^{\text{T}}=\left[\mathbf{\hat{h}}_1,\cdots, \mathbf{\hat{h}}_K\right]^{\text{T}}$. The error matrix lead to $\mathbf{H}^{\text{T}}=\mathbf{\hat{H}}^{\text{T}}+\mathbf{\tilde{H}}^{\text{T}}$. The error power induced by $\mathbf{\tilde{h}}_k$ is denoted by $\sigma_{e,k}^2=\mathbb{E}\left[\lVert\mathbf{\tilde{h}}_k\rVert^2\right], k=1,\cdots,K$. We also consider that the quality of the CSIT scales with the SNR, i.e. $\sigma_{e,k}^2=O\left(E_{tr}^{-\alpha}\right)$.

The received signal is given by
\begin{equation}
\mathbf{y}=\mathbf{H}^{\text{T}}\mathbf{P}\mathbf{v}+\mathbf{n}, \label{Receive signal complete}
\end{equation}
where $\mathbf{n}$ stands for the additive noise, which follows a complex Gaussian distribution $\mathbf{n}\sim\mathcal{CN}\left(\mathbf{0},\sigma_n^2\mathbf{I}\right)$. We consider that the SNR is equal to $E_{tr}/\sigma_n^2$. Furthermore we assume that the noise variance remains fixed so that the SNR scales with $E_{tr}$. From \eqref{Receive signal complete}, we obtain the received signal at each UE as expressed by
\begin{equation}
y_k=s_c\mathbf{h}_k^{\text{T}}\mathbf{p}_c+s_k\mathbf{h}_k^{\text{T}}\mathbf{p}_k^+\sum\limits_{\substack{i=1\\i\neq k}}^{K}s_i\mathbf{h}_k^{\text{T}}\mathbf{p}_i+n_k.
\end{equation}
The average power of the received signal at the $k$th user terminal is described by
\begin{equation}
\mathbb{E}\left[\lvert y_k\rvert^2\right]= \lvert \mathbf{h}_k^{\text{T}}\mathbf{p}_c\rvert^2+\lvert\mathbf{h}_k^{\text{T}}\mathbf{p}_k \rvert^2+\sum\limits_{\substack{i=1\\i\neq k}}^{K}\lvert \mathbf{h}_k^{\text{T}}\mathbf{p}_i\rvert^2+\sigma_n^2.
\end{equation}
The receiver decodes first the common symbol by considering the private symbols as noise and then performs SIC \cite{itic,DeLamare2008,jidf,mfsic,mbdf,tds,bfidd,aaidd,listmtc,1bitidd,detmtc,dynovs} to subtract the common symbol from the receive signal. Afterwards, the private symbols are decoded. Thus, the instantaneous sum rate of a system employing RS involves two different parts, the instantaneous common rate $R_c$ related to the common stream and the instantaneous sum-private rate associated to the private streams and given by $R_p=\sum_{k=1}^{K}R_k$, where $R_k$ denotes the $k$th user private rate. The common rate at the $k$th user is given by $R_{c,k}$. Assuming Gaussian codebooks we have 
\begin{align}
R_{c,k}&=\log_2\left(1+ \gamma_{c,k}\right), & R_{k}&=\log_2\left(1+ \gamma_{k}\right).\label{Rate Gaussian Codebook}
\end{align}   
where $\gamma_{c,k}$ represents the signal-to-interference-plus-noise ratio (SINR) at the $k$th user when decoding the common message and $\gamma_{k}$ denotes the SINR at the $k$th user when decoding its private message. We highlight that imperfect CSIT is considered in this work and therefore the instantaneous rates are not achievable. Consequently, we employ the ergodic sum rate (ESR), which is an achievable rate as detailed in \cite{Joudeh2016}, to measure the performance of the system. The ESR is  obtained from the average sum rate (ASR), which is a representation of the average performance of a given channel estimate taking into account the errors in the channel estimate. Similar to the instantaneous rate, the ASR consists of two parts. The average common rate is computed by $\bar{R}_{c,k}=\mathbb{E}_{\mathbf{\mathbf{H}^{\text{T}}|\hat{\mathbf{H}}^{\text{T}}}}\left[R_{c,k}|\hat{\mathbf{H}}^{\text{T}}\right]$, whereas the average sum private rate is given by $\bar{R}_{p}=\mathbb{E}_{\mathbf{H}^{\text{T}}|\hat{\mathbf{H}}^{\text{T}}}\left[R_{p}|\hat{\mathbf{H}}^{\text{T}}\right]$. The ergodic common rate can be obtained from $\mathbb{E}_{\hat{\mathbf{H}}^{\text{T}}}\left[\bar{R}_{c,k}\right]$ and the ergodic private rate is equal to $\mathbb{E}_{\hat{\mathbf{H}}^{\text{T}}}\left[\bar{R}_p\right]$. The ESR of the system can be expressed as follows:
\begin{equation}
S_r=\min_{k} \mathbb{E}_{\hat{\mathbf{H}}^{\text{T}}}\left[\bar{R}_{c,k}\right]+\mathbb{E}_{\hat{\mathbf{H}}^{\text{T}}}\left[\bar{R}_p\right].
\end{equation}
Note that the $\min$ operator is employed to guarantee that all users decode the common message.

\section{Proposed RS-MB-THP Scheme}
\label{sec:RS MB THP}

The THP was originally proposed in \cite{Tomlinson1971,Harashima1972} and extended to the MIMO environment in \cite{Fischer2002}. In the proposed RS-MB-THP scheme, one specific THP operates as the private precoder in $\mathbf{P}$ and requires the implementation of three filters: a feedback filter $\mathbf{B}$ which suppresses the interference caused by previous streams, a feedforward filter $\mathbf{F}$ which partially removes the MUI and a scaling matrix $\mathbf{C}$ which assigns a weight to each private stream. Depending on the position of the scaling matrix, we can define two different THP deployments. The centralized THP (cTHP) incorporates the matrix $\mathbf{C}$ at the receiver, whereas the decentralized THP (dTHP) implements the scaling matrix at the receivers\cite{Zu2014}. 
The three filters can be obtained by performing an LQ decomposition over the channel estimate which lead us to
\begin{equation}
 \hat{\mathbf{H}}^{\text{T}}=\hat{\mathbf{L}}\hat{\mathbf{Q}},
\end{equation}
 where $\hat{\mathbf{Q}}\in \mathbb{C}^{N_t\times N_t}$ is a unitary matrix. Then, we define the THP filters as follows:
 \begin{align}
\mathbf{F}&=\hat{\mathbf{Q}}^{H}\\
\mathbf{G}&=\text{diag}\left(\hat{l}_{1,1},\hat{l}_{2,2},\cdots,\hat{l}_{K,K}\right)\\
\mathbf{B}^{\left(\text{d}\right)}&=\mathbf{G}\hat{\mathbf{L}}~~ \text{and}~~ \mathbf{B}^{\left(\text{c}\right)}=\hat{\mathbf{L}}\mathbf{G} 
 \end{align}
 The received signal at the user terminal before the SIC operation is given by
 \begin{align}
 \mathbf{y}^{\left(\text{d}\right)}&=s_c\mathbf{H}^{\text{T}}\mathbf{p}_c+\beta \mathbf{H}^{\text{T}}\mathbf{F}\mathbf{v}+\mathbf{n},\\
 \mathbf{y}^{\left(\text{c}\right)}&=s_c\mathbf{H}^{\text{T}}\mathbf{p}_c+\beta\mathbf{H}^{\text{T}}\mathbf{F}\mathbf{C}\mathbf{v}+\mathbf{n},
 \end{align}
where $\beta$ is the power scaling factor and the vector $\mathbf{v}$ is obtained using the following recursion:
\begin{equation}
v_i=s_i-\sum_{j=1}^{i-1}b_{i,j}v_j.
\end{equation}
Note that the matrix $\mathbf{B}$ increases the average power of the transmit vector, which is reflected as a power loss\footnote{Remark that $v_1=s_1$ which lead us to $\sigma_{v_1}^2=\sigma_s^2$. However this statement does not hold for $v_i$ with $i\neq 1$ and we usually have that $\sigma_{v_i}\geq \sigma_s^2$.}. Thus, a modulo operation is introduced to reduce the effects of the power loss and keep the transmit vector inside the original modulation boundary \cite{Debels2018,Zu2014}. The modulo operation can be defined element-wise by
\begin{equation}
M\left(v_i\right)=v_i-\left \lfloor{\frac{\text{Re}\left(v_i\right)}{\lambda}+\frac{1}{2}}\right\rfloor\lambda-j\left\lfloor{\frac{\text{Im}\left(v_i\right)}{\lambda}+\frac{1}{2}}\right\rfloor\lambda,
\end{equation}
where $\lambda$ is a scalar that depends on the modulation alphabet (Considering a unitary symbol variance, some typical values of $\lambda$ are $2\sqrt{2}$ and $8\sqrt{10}$ for QPSK and 16-QAM respectively). Mathematically, the modulo operation corresponds to adding a perturbation vector $\mathbf{d}$ to the transmitted data. On the other hand, the result of feedback recursion can be also obtained by using the inverse of matrix $\mathbf{B}$, leading to the following transmit vector:
\begin{equation}
\mathbf{v}=\mathbf{B}^{-1}\left(\mathbf{s}+\mathbf{d}\right).
\end{equation}  
After decoding the common symbol and removing it from the receive signal, we obtain
\begin{align}
\mathbf{y}^{\left(\text{d}\right)}&=\beta\mathbf{C}\left(\mathbf{H}^{\text{T}}\mathbf{F}\mathbf{v}+\mathbf{n}\right),\\
\mathbf{y}^{\left(\text{c}\right)}&=\beta\mathbf{H}^{\text{T}}\mathbf{F}\mathbf{C}\mathbf{v}+\mathbf{n}.
\end{align}

\subsection{Power allocation}

Since we are using an RS approach, the power must be allocated to both the private streams and the common stream. Let us consider that the power of the common stream is given by $\lVert\mathbf{p}_c\rVert=\delta \sqrt{E_{tr}}$, where $\delta$ is the portion of $E_{tr}$ that is reserved for $s_c$. This parameter should be found so that the specific performance metric is maximized. Thus, the power for the private streams can be found by $\left(1-\delta\right)E_{tr}$ and, is uniformly allocated across streams. In this work, the parameter $\delta$ is chosen so that the ESR of the system is maximized. This can be expressed as follows:
\begin{align}
\delta_o=&\max_{\delta}\left(\min_{k\in \left[1,K\right]}\mathbb{E}_{\hat{\mathbf{H}}}\left[\bar{R}_{c,k}\left(\delta\right)\right]+\mathbb{E}_{\hat{\mathbf{H}}}\left[\bar{R}_p\left(\delta\right)\right]\right).\label{Power allocation SR}
\end{align}

Remark that the power constraint must be satisfied. For this purpose, a scaling factor $\beta$ is introduced into the private precoder, resulting in an average transmit power equal to $E_{tr}-\lVert\mathbf{p_c}\rVert^2$. Considering the THP as the private precoder, the value of $\beta$ is reduced to
\begin{align}
\beta^{\left(\textrm{cTHP}\right)}=&\sqrt{\frac{E_{tr}-\lVert \mathbf{p}_c\rVert^2}{\sum_{k=1}^{M}\hat{l}_{k,k}^2}} &
\beta^{\left(\textrm{dTHP}\right)}=&\sqrt{\frac{E_{tr}-\lVert \mathbf{p}_c\rVert^2}{M}}.
\end{align}

\subsection{Multi-Branch processing}

The (MB) technique was proposed in \cite{DeLamare2008} for decision feedback receivers and later extended to THP in \cite{Zu2014}. In this work, incorporate the MB concept into a RS-THP scheme and develop power allocation strategies for the proposed RS-THP-MB scheme. The MB technique employs multiple parallel branches to find the best symbol ordering for transmission. Let us suppose that $L_o$ different symbol ordering patterns originate $L_o$ transmit vectors candidates. The patterns are known to both, the transmitter and the receiver and each pattern is stored in a matrix $\mathbf{T}_{l}$ with $l\in 1,\cdots,L_o$ and $\hat{\mathbf{H}}^{\text{T}}_{\left(l\right)}=\mathbf{T}_l\hat{\mathbf{H}}^{\text{T}}$. Then, the transmitter finds the pattern that maximizes the desired metric, which in this work is the ESR performance. 

The first step of the MB approach is to rearrange the users according to several transmit patterns, which are described as follows:
\begin{align}
\mathbf{T}_{1}=&\mathbf{I}_K\\
\mathbf{T}_{i}=&\begin{bmatrix}
\mathbf{I}_{i-2} &\mathbf{0}_{i-2,K-i+2}\\
\mathbf{0}_{K-i+2,i-2} &\mathbf{\Pi}_i
\end{bmatrix}
,2\leq i\leq K.\label{User Order MB}
\end{align}
The matrix $\mathbf{\Pi}_i\in \mathbb{R}^{\left(K-i+2\right)\times\left(K-i+2\right)}$ is responsible for changing the order of the users. This  matrix contains ones in the reverse diagonal and zeros outside the reverse diagonal. 
Different form the conventional MU-MIMO, RS schemes include a common precoder besides the private precoders. The patterns created modify not only the private precoder but also the the common precoder. In this sense, the computation of the common precoder should be performed by taking into account the selected pattern. The best pattern for each channel estimate should be selected with the following criterion:

\begin{align}
\mathbf{T}_{\left(o\right)}&=\max_{l,\delta}\left(\min_{k}\bar{R}_{c,k}\left(\delta,\hat{\mathbf{H}}_{\left(l\right)}^{\text{T}} \right)+\bar{R}_p\left(\delta,\hat{\mathbf{H}}_{\left(l\right)}^{\text{T}} \right)\right).\label{Overall MB criterion}
\end{align}
Note that omitting the common rate in the selection criterion may lead us to select the wrong candidate. Once the best branch is selected, the channel estimate is rearranged accordingly, i.e., $\hat{\mathbf{H}}_{\left(o\right)}^{\text{T}}=\mathbf{T}_{\left(o\right)}\hat{\mathbf{H}}^{\text{T}}$. Then, the private and common precoders are computed.

Finding the optimal arrangement using \eqref{Overall MB criterion} results in a considerable high computational cost. To reduce the computational complexity other suboptimal approaches may be used. For this purpose, let us consider the fixed power allocation (FPA) criterion, which fixes $\delta$ to $\delta_f$, which is the $\delta$ that obtains the best ESR using $\hat{\mathbf{H}}^{\text{T}}$ without any rearrangement. Once the power allocation is set, the best branch for each channel estimate is selected according to
\begin{equation}
\mathbf{T}_{\left(o\right)}=\max_{l}\left(\min_{k}\bar{R}_{c,k}\left(\delta_f,\hat{\mathbf{H}}_{\left(l\right)}^{\text{T}} \right)+\bar{R}_p\left(\delta_f,\hat{\mathbf{H}}_{\left(l\right)}^{\text{T}} \right)\right).\label{MB criterion 2}
\end{equation}

A simplified approach, which we term fixed branch (FB) criterion, employs $\delta_f$ and chooses the branch that leads to the highest ESR as follows:  
\begin{align}
\mathbf{T}_{\left(o\right)}&=\nonumber\\
&\max_{l}\left(\min_{k}\mathbb{E}_{\hat{\mathbf{H}}_{\left(l\right)}^{\text{T}}}\left[\bar{R}_{c,k}\left(\hat{\mathbf{H}}^{\text{T}}_{\left(l\right)}\right)\right]+\mathbb{E}_{\hat{\mathbf{H}}^{\text{T}}_{\left(l\right)}}\left[\bar{R}_p\left(\hat{\mathbf{H}}^{\text{T}}_{\left(l\right)} \right)\right]\right).\label{MB criterion 3}
\end{align}
The branch is selected when the system is initialized, avoiding the search at each channel estimate. Then, the value of $\delta_f$ can be adapted using \eqref{Power allocation SR}. Since we consider the initial $\hat{\mathbf{H}}^{\text{T}}$ without symbol ordering as a possible solution of \eqref{Overall MB criterion},\eqref{MB criterion 2} and \eqref{MB criterion 3}, the proposed criteria perform at least as well as the conventional RS-THP. 

\section{Statistical analysis}
\label{sec:Statistical analysis}

Let us consider the $l$th branch of the proposed RS-MB-THP. The SINR expression for this branch when decoding the common stream at the $k$th user is given by
\begin{align}
\gamma^{\left(\text{d}\right)}_{c,l,k}&=\frac{\lvert\hat{\mathbf{h}}^{\left(l\right)^{\text{T}}}_k\mathbf{p}_c\rvert^2/\beta^{\left(\text{d}\right)}}{\lvert\hat{l}_{k,k}^{\left(l\right)}+\tilde{\mathbf{h}}_{k}^{\left(l\right)^{\text{T}}}\mathbf{p}_{k}^{\left(\text{d}\right)}\rvert^2+\sum\limits_{\substack{i=1\\ i\neq k}}^{K}\lvert\tilde{\mathbf{h}}_k^{\left(l\right)^{\text{T}}}\mathbf{p}_i^{\left(\text{d}\right)}\rvert^{2}+\sigma_n^2/\beta^{\left(\text{d}\right)}},\label{SINR dTHP common}\\
\gamma^{\left(\text{c}\right)}_{c,l,k}&=\frac{\lvert\hat{\mathbf{h}}^{\left(l\right)^{\text{T}}}_k\mathbf{p}_c\rvert^2/\beta^{\left(\text{c}\right)}}{\lvert 1+\tilde{\mathbf{h}}_{k}^{\left(l\right)^{\text{T}}}\mathbf{p}_{k}^{\left(\text{c}\right)}\rvert^2+\sum\limits_{\substack{i=1\\ i\neq k}}^{K}\lvert\tilde{\mathbf{h}}_k^{\left(l\right)^{\text{T}}}\mathbf{p}_i^{\left(\text{c}\right)}\rvert^{2}+\sigma_n^2/\beta^{\left(\text{c}\right)}}.\label{SINR cTHP common}
\end{align}

The associated SINR experienced by the $k$th user when decoding the private message with the $l$th branch can be computed by
\begin{align}
\gamma_{l,k}^{\left(\text{d}\right)}&=\frac{1}{\sum\limits_{\substack{i=1\\i\neq k}}^{K}\frac{1}{\hat{l}^2_{k,k}}\lvert\tilde{\mathbf{h}}_k^{\left(l\right)^{\text{T}}}\mathbf{p}_i^{\left(\text{d}\right)}\rvert^2+\frac{K \sigma_n^2}{\left(E_{tr}-\lVert\mathbf{p}_c\rVert^2\right)\hat{l}_{k,k}^2}},\label{SINR dTHP private}\\
\gamma_{l,k}^{\left(\text{c}\right)}&=\frac{1}{\sum\limits_{\substack{i=1\\i\neq k}}^{K}\lvert\tilde{\mathbf{h}}_k^{\left(l\right)^{\text{T}}}\mathbf{p}_i^{\left(\text{d}\right)}\rvert^2+\frac{\sigma_n^2\sum_j^K\left(1/\hat{l}_{j,j}\right)}{E_{tr}-\lVert\mathbf{p}_c\rVert^2}}.\label{SINR cTHP private}
\end{align}
Using \eqref{SINR cTHP common},\eqref{SINR dTHP common},\eqref{SINR cTHP private} and \eqref{SINR dTHP private} with \eqref{Rate Gaussian Codebook} and taking the expected value given $\hat{\mathbf{H}}_{\left(l\right)}$ we obtain the achievable rate of each user when employing the $l$th branch. Then, we select the best branch according the criteria established in the previous section.

\section{Simulations}
\label{sec:Simulations}

In this section, we assess the performance of the proposed RS-MB-THP and compare the results with the existing RS-THP and THP schemes \cite{Flores2018,Zu2014}. To this end, we consider the downlink of a MISO system where the BS is equipped with 8 transmit antennas. The BS establishes communication with 8 users geographically distributed. Moreover, we consider additive white Gaussian noise with $\sigma_n^2=1$. For the transmission, a flat fading Rayleigh channel is considered, which remains fix during the transmission of a packet. A total of 100 channel matrices was used to compute the ASR. On the other hand, the ESR was obtaind by averaging 100 channel estimates. An SVD was performed over the channel matrix ($\mathbf{H}=\mathbf{USV}$) and the common precoder is set to $\mathbf{p}_c=\mathbf{v}_1$. The optimal branch and the value of $\delta$ were obtained using \eqref{Overall MB criterion},\eqref{MB criterion 2} and \eqref{MB criterion 3}.

\begin{figure}[h]
\begin{centering}
\includegraphics[scale=0.35]{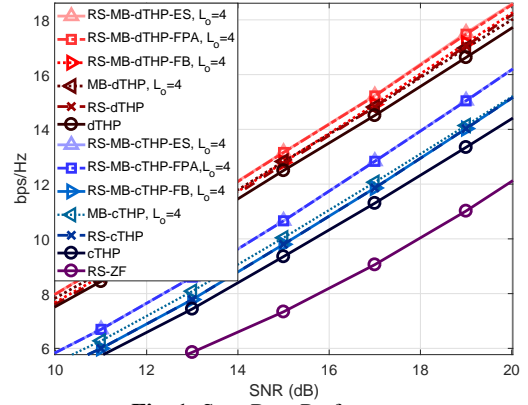}
\vspace{-1.5em}
\caption{Sum Rate Performance}
\label{Figure9}
\end{centering}
\end{figure}

In the first simulation, we consider that the quality of the channel estimate improves with the SNR. Specifically, we consider that the variance of the error is equal to $\sigma_e^2=0.95\left(E_{tr}^{-0.6}\right)$. Note that the variance of the error is inversely proportional to the SNR since the noise variance is fixed to one. We consider 4 different branches for both THP structures. Fig. \ref{Figure9} summarizes the overall sum rate performance, where the notation ES,FPA and FB stand for the exhaustive search, fixed power allocation and fixed branch criteria. The best performance is attained by the RS-MB-dTHP-ES as expected. Remark that the performance of the RS-MB-dTHP-FPA is slightly lower than the RS-MB-dTHP-ES but the computational complexity is reduced dramatically. The RS-MB-dTHP-FB technique obtains the worst performance among the proposed techniques but it also offers the lowest computational complexity.

\begin{figure}[h]
\begin{centering}
\includegraphics[scale=0.35]{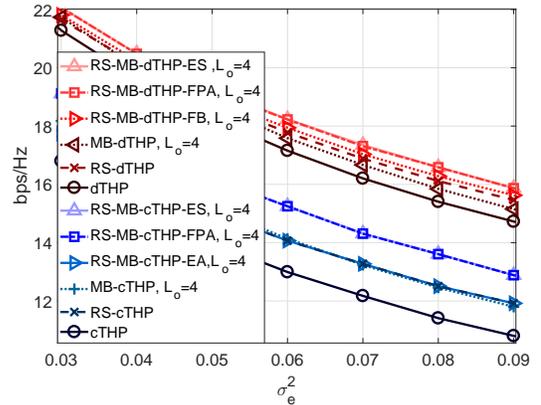}
\vspace{-1.5em}
\caption{Sum Rate Performance vs Error Variance}
\label{Figure7}
\end{centering}
\end{figure}

In the second example, we evaluate the robustness of the proposed algorithms by increasing the level in the CSIT uncertainty. We consider an SNR equal to 20dB. Fig. \ref{Figure7} shows the results obtained. The proposed strategies increase the robustness of the system across all error variances. The RS-MB-dTHP-ES is the most robust technique. Note that the RS-MB-dTHP-FPA is a suitable replacement for RS-MB-dTHP-ES with reduced computational complexity.

\section{Conclusions}
\label{sec:Conclusions}

In this work, we have proposed several RS-MB-THP techniques to perform symbol ordering and enhance the performance of RS-THP systems. In general, the proposed strategies exhibit greater benefits when dealing with CSIT imperfections. Results show that the proposed strategies achieve a better overall sum rate performance. Specifically, the RS-MB-dTHP-ES and RS-MB-dTHP-FPA obtain the best performance among the proposed strategies. Moreover, RS-MB-THP further enhances the robustness of the system even with a small number of branches, which demonstrates its effectiveness. 

%


\vfill\pagebreak

%

\bibliographystyle{IEEEbib}
\bibliography{THPStreamComb}

\begin{thebibliography}{10}

\bibitem{Li2010}
Q.~{Li}, G.~{Li}, W.~{Lee}, M.~{Lee}, D.~{Mazzarese}, B.~{Clerckx}, and
  Z.~{Li},
\newblock ``{MIMO} techniques in {WiMAX} and {LTE}: a feature overview,''
\newblock {\em IEEE Communications Magazine}, vol. 48, no. 5, pp. 86--92, May
  2010.

\bibitem{Jones2015}
V.~{Jones} and H.~{Sampath},
\newblock ``Emerging technologies for {WLAN},''
\newblock {\em IEEE Communications Magazine}, vol. 53, no. 3, pp. 141--149,
  Mar. 2015.

\bibitem{Parkvall2017}
S.~{Parkvall}, E.~{Dahlman}, A.~{Furuskar}, and M.~{Frenne},
\newblock ``{NR}: The new {5G} radio access technology,''
\newblock {\em IEEE Communications Standards Magazine}, vol. 1, no. 4, pp.
  24--30, Dec. 2017.

\bibitem{Shafi2017}
M.~{Shafi}, A.~F. {Molisch}, P.~J. {Smith}, T.~{Haustein}, P.~{Zhu}, P.~{De
  Silva}, F.~{Tufvesson}, A.~{Benjebbour}, and G.~{Wunder},
\newblock ``{5G}: A tutorial overview of standards, trials, challenges,
  deployment, and practice,''
\newblock {\em IEEE Journal on Selected Areas in Communications}, vol. 35, no.
  6, pp. 1201--1221, June 2017.

\bibitem{mmimo}
R.~C. de~Lamare,
\newblock ``Massive mimo systems: Signal processing challenges and future
  trends,''
\newblock {\em URSI Radio Science Bulletin}, vol. 2013, no. 347, pp. 8--20,
  2013.

\bibitem{wence}
Wence Zhang, Hong Ren, Cunhua Pan, Ming Chen, Rodrigo~C. de~Lamare, Bo~Du, and
  Jianxin Dai,
\newblock ``Large-scale antenna systems with ul/dl hardware mismatch:
  Achievable rates analysis and calibration,''
\newblock {\em IEEE Transactions on Communications}, vol. 63, no. 4, pp.
  1216--1229, 2015.

\bibitem{joham2005}
M.~{Joham}, W.~{Utschick}, and J.~A. {Nossek},
\newblock ``Linear transmit processing in {MIMO} communications systems,''
\newblock {\em IEEE Transactions on Signal Processing}, vol. 53, no. 8, pp.
  2700--2712, Aug. 2005.

\bibitem{Tomlinson1971}
M.~{Tomlinson},
\newblock ``New automatic equaliser employing modulo arithmetic,''
\newblock {\em Electronics Letters}, vol. 7, no. 5, pp. 138--139, Mar. 1971.

\bibitem{Harashima1972}
H.~{Harashima} and H.~{Miyakawa},
\newblock ``Matched-transmission technique for channels with intersymbol
  interference,''
\newblock {\em IEEE Transactions on Communications}, vol. 20, no. 4, pp.
  774--780, Aug. 1972.

\bibitem{Sung2009}
H.~{Sung}, S.~. {Lee}, and I.~{Lee},
\newblock ``Generalized channel inversion methods for multiuser mimo systems,''
\newblock {\em IEEE Transactions on Communications}, vol. 57, no. 11, pp.
  3489--3499, Nov. 2009.

\bibitem{gbd}
K.~Zu, R.~C. de~Lamare, and M.~Haardt,
\newblock ``Generalized design of low-complexity block diagonalization type
  precoding algorithms for multiuser mimo systems,''
\newblock {\em IEEE Transactions on Communications}, vol. 61, no. 10, pp.
  4232--4242, 2013.

\bibitem{wlbd}
W.~Zhang, R.~C. de~Lamare, C.~Pan, M.~Chen, J.~Dai, B.~Wu, and X.~Bao,
\newblock ``Widely linear precoding for large-scale mimo with iqi: Algorithms
  and performance analysis,''
\newblock {\em IEEE Transactions on Wireless Communications}, vol. 16, no. 5,
  pp. 3298--3312, 2017.

\bibitem{Zu2014}
K.~{Zu}, R.~C. {de Lamare}, and M.~{Haardt},
\newblock ``Multi-branch tomlinson-harashima precoding design for {MU-MIMO}
  systems: Theory and algorithms,''
\newblock {\em IEEE Transactions on Communications}, vol. 62, no. 3, pp.
  939--951, Mar. 2014.

\bibitem{rthp}
L.~Zhang, Y.~Cai, R.~C. de~Lamare, and M.~Zhao,
\newblock ``Robust multibranch tomlinson–harashima precoding design in
  amplify-and-forward mimo relay systems,''
\newblock {\em IEEE Transactions on Communications}, vol. 62, no. 10, pp.
  3476--3490, 2014.

\bibitem{rmmse}
Y.~Cai, R.~C. de~Lamare, L.-L. Yang, and M.~Zhao,
\newblock ``Robust mmse precoding based on switched relaying and side
  information for multiuser mimo relay systems,''
\newblock {\em IEEE Transactions on Vehicular Technology}, vol. 64, no. 12, pp.
  5677--5687, 2015.

\bibitem{bbprec}
L.~T.~N. Landau and R.~C. de~Lamare,
\newblock ``Branch-and-bound precoding for multiuser mimo systems with 1-bit
  quantization,''
\newblock {\em IEEE Wireless Communications Letters}, vol. 6, no. 6, pp.
  770--773, 2017.

\bibitem{baplnc}
Jiaqi Gu, Rodrigo~C. de~Lamare, and Mario Huemer,
\newblock ``Buffer-aided physical-layer network coding with optimal linear code
  designs for cooperative networks,''
\newblock {\em IEEE Transactions on Communications}, vol. 66, no. 6, pp.
  2560--2575, 2018.

\bibitem{lrcc}
Hang Ruan and Rodrigo~C. de~Lamare,
\newblock ``Distributed robust beamforming based on low-rank and
  cross-correlation techniques: Design and analysis,''
\newblock {\em IEEE Transactions on Signal Processing}, vol. 67, no. 24, pp.
  6411--6423, 2019.

\bibitem{mwmimo}
Flavio~L. Duarte and Rodrigo~C. de~Lamare,
\newblock ``Cloud-driven multi-way multiple-antenna relay systems: Joint
  detection, best-user-link selection and analysis,''
\newblock {\em IEEE Transactions on Communications}, vol. 68, no. 6, pp.
  3342--3354, 2020.

\bibitem{plszf}
,''
\newblock .

\bibitem{rmmsecf}
Victoria M.~T.~Palhares, Andre Flores, and Rodrigo~C. De~Lamare,
\newblock ``Robust mmse precoding and power allocation for cell-free massive
  mimo systems,''
\newblock {\em IEEE Transactions on Vehicular Technology}, pp. 1--1, 2021.

\bibitem{Clerckx2016}
B.~{Clerckx}, H.~{Joudeh}, C.~{Hao}, M.~{Dai}, and B.~{Rassouli},
\newblock ``Rate splitting for {MIMO} wireless networks: a promising
  {PHY}-layer strategy for {LTE} evolution,''
\newblock {\em IEEE Communications Magazine}, vol. 54, no. 5, pp. 98--105, May
  2016.

\bibitem{Mao2018}
Y.~{Mao}, B.~{Clerckx}, and V.~{Li},
\newblock ``Rate-splitting multiple access for downlink communication systems:
  Bridging, generalizing and outperforming {SDMA} and {NOMA},''
\newblock {\em EURASIP Journal on Wireless Communications and Networking}, vol.
  2018, no. 1, pp. 133, 2018.

\bibitem{Mao2020}
Y.~{Mao} and B.~{Clerckx},
\newblock ``Beyond dirty paper coding for multi-antenna broadcast channel with
  partial {CSIT}: A rate-splitting approach,''
\newblock {\em accepted to IEEE Transactions on Communications}, 2020.

\bibitem{Joudeh2016}
H.~{Joudeh} and B.~{Clerckx},
\newblock ``Sum-rate maximization for linearly precoded downlink multiuser
  {MISO} systems with partial {CSIT}: A rate-splitting approach,''
\newblock {\em IEEE Transactions on Communications}, vol. 64, no. 11, pp.
  4847--4861, Nov. 2016.

\bibitem{Hao2015}
C.~{Hao}, Y.~{Wu}, and B.~{Clerckx},
\newblock ``Rate analysis of two-receiver {MISO} broadcast channel with finite
  rate feedback: A rate-splitting approach,''
\newblock {\em IEEE Transactions on Communications}, vol. 63, no. 9, pp.
  3232--3246, Sept. 2015.

\bibitem{Lu2018}
G.~{Lu}, L.~{Li}, H.~{Tian}, and F.~{Qian},
\newblock ``{MMSE}-based precoding for rate splitting systems with finite
  feedback,''
\newblock {\em IEEE Communications Letters}, vol. 22, no. 3, pp. 642--645, Mar.
  2018.

\bibitem{Flores2018}
A.~R. {Flores}, B.~{Clerckx}, and R.~C. {de Lamare},
\newblock ``Tomlinson-harashima precoded rate-splitting for multiuser
  multiple-antenna systems,''
\newblock in {\em 2018 15th International Symposium on Wireless Communication
  Systems (ISWCS)}, Lisbon, 2018, pp. 1--6.

\bibitem{rsthp}
A.~R. Flores, R.~C. De~Lamare, and B.~Clerckx,
\newblock ``Tomlinson-harashima precoded rate-splitting with stream combiners
  for mu-mimo systems,''
\newblock {\em IEEE Transactions on Communications}, pp. 1--1, 2021.

\bibitem{Vu2007}
M.~{Vu} and A.~{Paulraj},
\newblock ``{MIMO} wireless linear precoding,''
\newblock {\em IEEE Signal Processing Magazine}, vol. 24, no. 5, pp. 86--105,
  Sept. 2007.

\bibitem{Love2008}
D.~J. {Love}, R.~W. {Heath}, V.~K. {N. Lau}, D.~{Gesbert}, B.~D. {Rao}, and
  M.~{Andrews},
\newblock ``An overview of limited feedback in wireless communication
  systems,''
\newblock {\em IEEE Journal on Selected Areas in Communications}, vol. 26, no.
  8, pp. 1341--1365, Oct. 2008.

\bibitem{itic}
Rodrigo~C. De~Lamare, Raimundo Sampaio-Neto, and Are Hjorungnes,
\newblock ``Joint iterative interference cancellation and parameter estimation
  for cdma systems,''
\newblock {\em IEEE Communications Letters}, vol. 11, no. 12, pp. 916--918,
  2007.

\bibitem{DeLamare2008}
R.~C. {De Lamare} and R.~{Sampaio-Neto},
\newblock ``Minimum mean-squared error iterative successive parallel arbitrated
  decision feedback detectors for {DS-CDMA} systems,''
\newblock {\em IEEE Transactions on Communications}, vol. 56, no. 5, pp.
  778--789, May 2008.

\bibitem{jidf}
Rodrigo~C. de~Lamare and Raimundo Sampaio-Neto,
\newblock ``Adaptive reduced-rank processing based on joint and iterative
  interpolation, decimation, and filtering,''
\newblock {\em IEEE Transactions on Signal Processing}, vol. 57, no. 7, pp.
  2503--2514, 2009.

\bibitem{mfsic}
Peng Li, Rodrigo~C. de~Lamare, and Rui Fa,
\newblock ``Multiple feedback successive interference cancellation detection
  for multiuser mimo systems,''
\newblock {\em IEEE Transactions on Wireless Communications}, vol. 10, no. 8,
  pp. 2434--2439, 2011.

\bibitem{mbdf}
Rodrigo~C. de~Lamare,
\newblock ``Adaptive and iterative multi-branch mmse decision feedback
  detection algorithms for multi-antenna systems,''
\newblock {\em IEEE Transactions on Wireless Communications}, vol. 12, no. 10,
  pp. 5294--5308, 2013.

\bibitem{tds}
Patrick Clarke and Rodrigo~C. de~Lamare,
\newblock ``Transmit diversity and relay selection algorithms for multirelay
  cooperative mimo systems,''
\newblock {\em IEEE Transactions on Vehicular Technology}, vol. 61, no. 3, pp.
  1084--1098, 2012.

\bibitem{bfidd}
Andre G.~D. Uchoa, Cornelius~T. Healy, and Rodrigo~C. de~Lamare,
\newblock ``Iterative detection and decoding algorithms for mimo systems in
  block-fading channels using ldpc codes,''
\newblock {\em IEEE Transactions on Vehicular Technology}, vol. 65, no. 4, pp.
  2735--2741, 2016.

\bibitem{aaidd}
Roberto~B. Di~Renna and Rodrigo~C. de~Lamare,
\newblock ``Adaptive activity-aware iterative detection for massive
  machine-type communications,''
\newblock {\em IEEE Wireless Communications Letters}, vol. 8, no. 6, pp.
  1631--1634, 2019.

\bibitem{listmtc}
Roberto~B. Di~Renna and Rodrigo~C. de~Lamare,
\newblock ``Iterative list detection and decoding for massive machine-type
  communications,''
\newblock {\em IEEE Transactions on Communications}, vol. 68, no. 10, pp.
  6276--6288, 2020.

\bibitem{1bitidd}
Zhichao Shao, Rodrigo~C. de~Lamare, and Lukas T.~N. Landau,
\newblock ``Iterative detection and decoding for large-scale multiple-antenna
  systems with 1-bit adcs,''
\newblock {\em IEEE Wireless Communications Letters}, vol. 7, no. 3, pp.
  476--479, 2018.

\bibitem{detmtc}
Roberto~B. Di~Renna, Carsten Bockelmann, Rodrigo~C. de~Lamare, and Armin
  Dekorsy,
\newblock ``Detection techniques for massive machine-type communications:
  Challenges and solutions,''
\newblock {\em IEEE Access}, vol. 8, pp. 180928--180954, 2020.

\bibitem{dynovs}
Zhichao Shao, Lukas T.~N. Landau, and Rodrigo~C. de~Lamare,
\newblock ``Dynamic oversampling for 1-bit adcs in large-scale multiple-antenna
  systems,''
\newblock {\em IEEE Transactions on Communications}, vol. 69, no. 5, pp.
  3423--3435, 2021.

\bibitem{Fischer2002}
R.~F.~H. {Fischer}, C.~{Windpassinger}, A.~{Lampe}, and J.~B. {Huber},
\newblock ``Space-time transmission using {Tomlinson-Harashima} precoding,''
\newblock in {\em 2002 ITG Conf. on Source and Channel Coding}, 2002, pp.
  139--147.

\bibitem{Debels2018}
E.~{Debels} and M.~{Moeneclaey},
\newblock ``{SNR} maximization and modulo loss reduction for
  {Tomlinson-Harashima} precoding,''
\newblock {\em EURASIP Journal on Wireless Communications and Networking}, vol.
  2018, no. 257, 2018.

\end{thebibliography}
\end{document}